# On the Thermodynamic Origin of the Quantum Potential

Gerhard Grössing,

*Austrian Institute for Nonlinear Studies,*

Akademiehof,

Friedrichstrasse 10, A-1010 Vienna, Austria

e-mail: ains@chello.at

**Abstract:** The quantum potential is shown to result from the presence of a subtle thermal vacuum energy distributed across the whole domain of an experimental setup. Explicitly, its form is demonstrated to be exactly identical to the heat distribution derived from the defining equation for classical diffusion-wave fields. For a single free particle path, this thermal energy does not significantly affect particle motion. However, in between different paths, or at interfaces, the accumulation-depletion law for diffusion waves provides an immediate new understanding of the quantum potential's main features.

"A theory is the more impressive the greater the simplicity of its premises, the more varied the kinds of things that it relates and the more extended the area of its applicability. Therefore classical thermodynamics has made a deep impression on me. It is the only physical theory of universal content which I am convinced, within the areas of the applicability of its basic concepts, will never be overthrown." - Albert Einstein (1949)

## 1. Introduction: Diffusion Waves

In the de Broglie – Bohm interpretation of the quantum mechanical formalism, a very special role is attributed to the quantum potential

$$U(\mathbf{x},t) = -\frac{\hbar^2}{2m}\frac{\nabla^2 R(\mathbf{x},t)}{R(\mathbf{x},t)}, \qquad (1.1)$$

with $R(\mathbf{x},t) = |\psi(\mathbf{x},t)|$, and $\psi(\mathbf{x},t)$ being the quantum mechanical wavefunction. The particular qualitative features of $U$ are considered to represent a radical departure from classical Newtonian physics. For example, Bohm and Hiley [1] point out that $U$ is not changed when the field $\psi$ is multiplied by an arbitrary constant (i.e., as $\psi$ appears both in the numerator and the denominator of $U$), and they comment on this as follows:

> "This means that the effect of the quantum potential is independent of the strength (i.e., the intensity) of the quantum field but depends only on its *form*. By contrast, classical waves, which act mechanically (i.e., to transfer energy and momentum, for example, to push a floating object), always produce effects that are more or less proportional to the strength of the wave." [1, p.31]

Moreover, after illustrating the new features of particle movement with the analogy of a ship on automatic pilot guided by external radio signals, the authors argue that

> "particles moving in empty space under the action of no classical forces need not travel uniformly in straight lines. (…) Moreover, since the effect of the wave does not necessarily fall off with the distance, even remote features of the environment can profoundly affect the movement." [1, p. 32]

Up to today, this is the state-of-the-art as far as explanations of the action of the quantum potential are concerned: the spatial "form" of the wave function is made responsible for radical deviations from classical physics as well as for the possibility of nonlocal correlations.

However, in this paper it is intended to provide a more physical, as opposed to a mere formal, explanation of the action of the quantum potential. It will turn out that in doing so, no departure from classical physics will be necessary. On the contrary, it is additional features from classical physics, which in the discussion of quantum mechanics so far have remained largely unnoticed, that will provide a more complete picture of how the quantum potential affects particle movement. The said additional features are those of classical diffusion waves, and thus, before we begin a discussion of quantum mechanics, some introductory remarks on them seem appropriate.

Mathematically, diffusion waves are characterized by the peculiarity that the time derivative in their defining equation is only of first order. This brings them into stark contrast to most other well-known wave phenomena, where linear wave equations contain a nonzero second-order time derivative (and which, one may add, Bohm and Hiley seemed to have had in mind when discussing classical waves). It is the latter property of hyperbolic equations which provides families of solutions in terms of forward and backward waves propagating in space and thus the familiar features we normally associate with waves, i.e., wavefront propagation, interference, or reflections at interfaces, for example. In contrast, no such solutions generally exist for the parabolic equations of diffusion waves, which are wave-like disturbances characterized by coherent, and always driven, oscillations of diffusing energy or particles. They do not exhibit square-law behavior, but show an infinite speed of propagation of disturbances along entire domains.

More particularly, diffusion waves arise when the classical diffusion equation (or, the heat conduction equation, respectively) is coupled to an oscillatory force function [2],

$$\nabla^2 \Psi(\mathbf{r},t) - \frac{1}{D}\frac{\partial}{\partial t}\Psi(\mathbf{r},t) = q(\mathbf{r})e^{i\omega t}, \qquad (1.2)$$

where the driving force $q(\mathbf{r})e^{i\omega t}$ generates oscillatory solutions for the field's wave function

$$\Psi(\mathbf{r},t) = \Phi(\mathbf{r},\omega)e^{i\omega t}. \qquad (1.3)$$

As we shall see, $\Psi(\mathbf{r},t)$ can be an oscillating heat flow, thus referring to thermal waves, where an extra term often added in (1.2), i.e., the square of the characteristic decay length of the diffusion waves, is usually vanishing identically and therefore being omitted here.

Applying a Fourier transform on (1.2), one obtains a "pseudo-wave Helmholtz equation" [2], i.e.,

$$\nabla^2 \Phi(\mathbf{r},\omega) - \kappa^2(\mathbf{r},\omega)\Phi(\mathbf{r},\omega) = Q(\mathbf{r},\omega), \qquad (1.4)$$

where $\kappa(\mathbf{r},\omega)$ is the complex-valued diffusion wavenumber, which embodies much of the physics of diffusion waves. With the thermal diffusivity $D$ and the diffusion length

$$L(\omega) = \sqrt{\frac{2D}{\omega}}, \qquad (1.5)$$

the expression for the diffusion wavenumber of thermal waves becomes

$$\kappa = \frac{1+i}{L(\omega)}, \qquad (1.6)$$

and thus

$$\kappa^2 = \frac{2i}{L^2} = \frac{i\omega}{D}. \qquad (1.7)$$

The transport of diffusion waves is usually given by Fick's law of diffusion. Instead of a square law, this is a linear law describing the transport via spatial diffusion gradients. If a source oscillating at an angular frequency $\omega$ in a medium of diffusivity $D$ generates an energy density or particle concentration $\rho(\mathbf{r},\omega)$, the resulting current density $\mathbf{J}(\mathbf{r},\omega)$ is given by Fick's first law,

$$\mathbf{J}(\mathbf{r},\omega) = -D\nabla\rho(\mathbf{r},\omega), \qquad (1.8)$$

and combination with the continuity equation $\dot{\rho} = -\nabla \cdot \mathbf{J}$ provides Fick's second law,

$$\frac{\partial \rho}{\partial t} = D\nabla^2 \rho. \qquad (1.9)$$

Andreas Mandelis has pointed out a general particularity of diffusion waves, which is very important also with respect to our discussion of the quantum potential:

> "The simple fact that the diffusion wave field propagates according to a linear law affects the waves' behaviour at interfaces. When they encounter an interface, diffusion waves obey an accumulationdepletion law, rather than the reflection-refraction law of normal waves. Because detecting diffusion waves almost always involves the waves crossing an interface of some sort, and because diffusion waves, being heavily damped, don't travel very far, their behaviour at interfaces is of great practical importance." [2]

Moreover, and most importantly, with diffusion waves spatial coherence can be created out of random ensembles of diffusive energy. In this context it is interesting that equations like (1.4) yield "the physical artefact of infinite speed of field propagation, though with vanishingly small amplitude, at remote locations away from the source. (…) Because propagation is instantaneous, the equations yield no travelling waves, no wavefronts, and no phase velocity. Rather, the entire domain 'breathes' in phase with the oscillating source. In the world of diffusion waves, there are only spatially correlated phase lags controlled by the diffusion length." [2] Naturally, if any phenomena from classical physics should be helpful at all in this regard, these features make diffusion waves particularly amenable for modelling quantum mechanical nonlocality. In what follows, it shall be shown that actually they are at the very core of a deeper understanding of the quantum potential.

This paper is structured as follows. In Section 2, the motivation for introducing diffusion-wave fields in the discussion on the foundations of quantum mechanics is explained by giving a short summary of the recently published derivation of the Schrödinger equation from nonequilibrium thermodynamics [3]. In an extension to ref.

[3], it is here given in terms of the many-particle Schrödinger equation. Within the thus established framework, it is shown in Section 3 that the expression for the quantum potential is identical to a form of the heat equation, or of the defining equation for diffusion-wave fields, respectively. In Section 4, it is shown with some examples how the form of the quantum potential is actually a representation of the accumulation-depletion law for diffusion wave fields. Finally, in Section 5 some conclusions and future perspectives are discussed.

**2. Motivation: Exact Schrödinger equation derived from nonequilibrium thermodynamics**

In reference [3], the fact that each particle of nature is attributed an energy $E = \hbar\omega$ is considered as central for the understanding of quantum theory. As oscillations, characterized by some typical angular frequency $\omega$, are today considered as properties of off-equilibrium steady-state systems, one can generally assume that "particles" are actually dissipative systems maintained in a nonequilibrium steady-state by a permanent throughput of energy, or heat flow, respectively. The latter must be expressed by some form of kinetic energy, which is not identical with the kinetic energy of the particle, but an additional (external) contribution to it. One can thus write down the energy of the total system, i.e., the particle as the "system of interest" and the heat flow as the particle's thermal context (*Assumption 1*):

$$E_{\text{tot}} = \hbar\omega + \frac{(\delta p)^2}{2m}, \qquad (2.1)$$

where $\delta p$ is the additional, fluctuating momentum component of the particle of mass $m$.

Moreover, as has been suggested since the beginnings of quantum theory, the particle's environment may be considered such as to provide detection probability distributions which can be modelled by wave-like intensity distributions. Thus, the detection probability density $P(x,t)$ is considered to coincide with a classical wave's intensity $I(x,t) = R^2(x,t)$, with $R(x,t)$ being the wave's real-valued amplitude (*Assumption 2*):

$$P(\mathbf{x},t) = R^2(\mathbf{x},t), \text{ with normalization } \int P \, d^n x = 1 . \qquad (2.2)$$

In the present paper, there is no room to cover the fascinating results which throughout recent years have been accumulated in the field of nonequilibrium thermodynamics. (For an excellent review, see, e.g., ref. [4].) In ref. [3], we proposed to merge some results of said field with classical wave mechanics in such a way that the many microscopic degrees of freedom associated with the hypothesized subquantum medium can be recast into the more "macroscopic" properties that characterize the wave-like behaviour on the quantum level. Thus, the relevant description of the "system of interest" no longer needs the full phase space information of all "microscopic" entities, but needs only the "emergent" particle coordinates.

As we consider a particle as being surrounded by a "heat bath", i.e., a reservoir that is very large compared to the small dissipative system, one can safely assume that the momentum distribution in this region is given by the usual Maxwell-Boltzmann distribution – a wide-spread characteristic for similar systems, despite the fact that we deal with nonequilibrium thermodynamics. This corresponds to a "thermostatic" regulation of the reservoir's temperature, which is equivalent to the statement that the energy lost to the thermostat can be regarded as heat. Thus, without further delving into the matter more deeply here, one arrives via a corresponding *proposition of emergence* [3] at the equilibrium-type probability (density) ratio (*Assumption 3*)

$$\frac{P(\mathbf{x},t)}{P(\mathbf{x},0)} = e^{-\frac{\Delta Q}{kT}}, \qquad (2.3)$$

with $k$ being Boltzmann's constant, $T$ the reservoir temperature, and $\Delta Q$ the heat that is exchanged between the particle and its environment.

Equations (2.1), (2.2), and (2.3) are the only assumptions necessary to derive the Schrödinger equation from (modern) classical mechanics. In fact, we need only two additional well-known observations in order to achieve that goal. The first is given by Boltzmann's formula for the slow transformation of a periodic motion (with period $\tau = 2\pi/\omega$) upon application of a heat transfer $\Delta Q$. With the action function $S = \int (E_{\text{kin}} - V) dt$, the relation between heat and action is given as

$$\Delta Q = 2\omega \delta S = 2\omega \left[ \delta S(t) - \delta S(0) \right]. \qquad (2.4)$$

Finally, as the kinetic energy of the thermostat is given by $kT/2$ per degree of freedom, and as the average kinetic energy of the oscillator is equal to half of the total energy $E = \hbar\omega$, the requirement that the average kinetic energy of the thermostat equals the average kinetic energy of the oscillator reads, for each degree of freedom, as

$$\frac{kT}{2} = \frac{\hbar\omega}{2}. \qquad (2.5)$$

Now, combining the Equs. (2.3), (2.4), and (2.5), one obtains

$$P(\mathbf{x},t) = P(\mathbf{x},0) e^{-\frac{2}{\hbar}\left[\delta S(\mathbf{x},t) - \delta S(\mathbf{x},0)\right]}, \qquad (2.6)$$

from which follows the expression for the momentum fluctuation $\delta \mathbf{p}$ of (2.1) as

$$\delta \mathbf{p}(\mathbf{x},t) = \nabla(\delta S(\mathbf{x},t)) = -\frac{\hbar}{2}\frac{\nabla P(\mathbf{x},t)}{P(\mathbf{x},t)}. \tag{2.7}$$

This, then, provides the additional kinetic energy term for one particle as

$$\delta E_{\text{kin}} = \frac{1}{2m}\nabla(\delta S)\cdot\nabla(\delta S) = \frac{1}{2m}\left(\frac{\hbar}{2}\frac{\nabla P}{P}\right)^2. \tag{2.8}$$

Thus, writing down a classical action integral for $n$ particles, including this new term for each of them, yields (with external potential $V$)

$$A = \int L\, d^n x dt = \int P\left[\frac{\partial S}{\partial t} + \sum_{i=1}^{n}\frac{1}{2m_i}\nabla_i S\cdot\nabla_i S + \sum_{i=1}^{n}\frac{1}{2m_i}\left(\frac{\hbar}{2}\frac{\nabla_i P}{P}\right)^2 + V\right] d^n x dt, \tag{2.9}$$

where $P = P(\mathbf{x}_1,\mathbf{x}_2,...,\mathbf{x}_n,t)$. Introducing the "Madelung transformation"

$$\psi = R e^{\frac{i}{\hbar}S}, \tag{2.10}$$

where $R = \sqrt{P}$ as in (2.2), one has, with bars denoting averages,

$$\overline{\left|\frac{\nabla_i\psi}{\psi}\right|^2} := \int d^n x dt \left|\frac{\nabla_i\psi}{\psi}\right|^2 = \overline{\left(\frac{1}{2}\frac{\nabla_i P}{P}\right)^2} + \overline{\left(\frac{\nabla_i S}{\hbar}\right)^2}, \tag{2.11}$$

and one can rewrite (2.9) as

$$A = \int L dt = \int d^n x dt \left[|\psi|^2\left(\frac{\partial S}{\partial t}+V\right) + \sum_{i=1}^{n}\frac{\hbar^2}{2m_i}|\nabla_i\psi|^2\right]. \tag{2.12}$$

Thus, with the identity $|\psi|^2\frac{\partial S}{\partial t} = -\frac{i\hbar}{2}(\psi^*\dot\psi - \dot\psi^*\psi)$, one obtains the familiar Lagrange density

$$L = -\frac{i\hbar}{2}(\psi^*\dot\psi - \dot\psi^*\psi) + \sum_{i=1}^{n}\frac{\hbar^2}{2m_i}\nabla_i\psi\cdot\nabla_i\psi^* + V\psi^*\psi, \tag{2.13}$$

from which by the usual procedures one arrives at the $n$–particle Schrödinger equation

$$i\hbar \frac{\partial \psi}{\partial t} = \left( -\sum_{i=1}^{n} \frac{\hbar^2}{2m_i} \nabla_i^2 + V \right) \psi. \tag{2.14}$$

Note also that from (2.9) one obtains upon variation in $P$ the modified Hamilton-Jacobi equation familiar from the de Broglie-Bohm interpretation, i.e.,

$$\frac{\partial S}{\partial t} + \sum_{i=1}^{n} \frac{(\nabla_i S)^2}{2m_i} + V(\mathbf{x}_1, \mathbf{x}_2, ..., \mathbf{x}_n, t) + U(\mathbf{x}_1, \mathbf{x}_2, ..., \mathbf{x}_n, t) = 0, \tag{2.15}$$

where $U$ is known as the "quantum potential"

$$U(\mathbf{x}_1, \mathbf{x}_2, ..., \mathbf{x}_n, t) = \sum_{i=1}^{n} \frac{\hbar^2}{4m_i} \left[ \frac{1}{2} \left( \frac{\nabla_i P}{P} \right)^2 - \frac{\nabla_i^2 P}{P} \right] = -\sum_{i=1}^{n} \frac{\hbar^2}{2m_i} \frac{\nabla_i^2 R}{R}. \tag{2.16}$$

Moreover, with the definitions

$$\mathbf{u}_i := \frac{\delta \mathbf{p}_i}{m_i} = -\frac{\hbar}{2m_i} \frac{\nabla_i P}{P} \text{ and } \mathbf{k}_{\mathbf{u}i} = -\frac{1}{2} \frac{\nabla_i P}{P} = -\frac{\nabla_i R}{R}, \tag{2.17}$$

one can rewrite $U$ as

$$U = \sum_{i=1}^{n} \left[ \frac{m_i \mathbf{u}_i \cdot \mathbf{u}_i}{2} - \frac{\hbar}{2} (\nabla_i \cdot \mathbf{u}_i) \right] = \sum_{i=1}^{n} \left[ \frac{\hbar^2}{2m_i} (\mathbf{k}_{\mathbf{u}i} \cdot \mathbf{k}_{\mathbf{u}i} - \nabla_i \cdot \mathbf{k}_{\mathbf{u}i}) \right]. \tag{2.18}$$

(Note that there is a sign error in Equ. (3.2.29) of ref. [3].) However, as was already pointed out in ref. [3], with the aid of (2.4) and (2.6), $\mathbf{u}_i$ can also be written as

$$\mathbf{u}_i = \frac{1}{2\omega_i m_i} \nabla_i Q, \tag{2.19}$$

which thus explicitly shows its dependence on the spatial behavior of the heat flow $\delta Q$. Insertion of (2.19) into (2.18) then provides the thermodynamic formulation of the quantum potential as

$$U = \sum_{i=1}^{n} \frac{\hbar^2}{4m_i} \left[ \frac{1}{2} \left( \frac{\nabla_i Q}{\hbar \omega_i} \right)^2 - \frac{\nabla_i^2 Q}{\hbar \omega_i} \right]. \tag{2.20}$$

## 3. The thermodynamic origin of the quantum potential

The energetic scenario of a steady-state oscillator in nonequilibrium thermodynamics is given by a throughput of heat, i.e., a kinetic energy at the subquantum level providing a) the necessary energy to maintain a constant oscillation frequency $\omega$, and b) some excess kinetic energy resulting in a fluctuating momentum contribution $\delta \mathbf{p}$ to the momentum $\mathbf{p}$ of the particle. From a perspective out of everyday life, one can compare this to the situation of some small convex half-sphere, say, lying on a flat vibrating membrane. Due to resonance, the half-sphere will oscillate with the same frequency as the membrane, but if the energy of the membrane's vibration is higher than that required for the half-sphere to co-oscillate, the latter will start to perform an irregular motion, thus reflecting minute irregularities in the membrane (or the half-sphere itself) such as to amplify them in a momentum fluctuation. However, there is one more element in the energy scenario that is important. In our everyday life example, it is the friction between the half-sphere and the membrane, which causes the half-sphere to dissipate heat energy into its environment.

Very similarly, the steady-state resonator representing a "particle" in a thermodynamic environment will not only receive kinetic energy from it, but, in order to balance the stochastic influence of the buffeting momentum fluctuations, it will also dissipate heat into the environment. In fact, the "vacuum fluctuation theorem" (VFT) introduced in ref. [3] proposes, as all fluctuation theorems, that the larger the energy fluctuation of the oscillating "system of interest" is, the higher is the probability that heat will be dissipated into the environment rather than be absorbed. Also, Bohm and Hiley [1] demand in their review of stochastic hidden variable models that, generally, to maintain an equilibrium density distribution like the one given by $P(x,t)$ under random processes, the latter *must* be complemented by a balancing movement. The corresponding balancing velocity is called, referring to the same expression in Einstein's work on Brownian motion, the "osmotic velocity". If we remind ourselves of the stochastic "forward" movement in our model, i.e., $\delta\mathbf{p}/m = \mathbf{u}$, or the current $\mathbf{J} = P\mathbf{u}$, respectively, this will have to be balanced by the osmotic velocity $-\mathbf{u}$, or $\mathbf{J} = -P\mathbf{u}$, respectively. (For simplicity, we restrict ourselves in the following discussion to the one-particle case, thus omitting the indices $i$ again.)

Inserting (2.17) into the definition of the "forward" diffusive current $\mathbf{J}$, and recalling the diffusivity $D = \hbar/2m$, one has

$$\mathbf{J} = P\mathbf{u} = -D\nabla P, \tag{3.1}$$

which, when combined with the continuity equation $\dot{P} = -\nabla \cdot \mathbf{J}$, becomes

$$\frac{\partial P}{\partial t} = D\nabla^2 P. \tag{3.2}$$

Equs. (3.1) and (3.2) are, in complete analogy to (1.8) and (1.9), the first and second of Fick's laws of diffusion, respectively, and $\mathbf{J}$ is the diffusion current.

So, whereas in ref. [3] we concentrated on that part of the energy throughput maintaining the particle's frequency $\omega$ that led to an additional momentum contribution $\delta\mathbf{p}$ from the environment to be absorbed by the particle, we now are going to focus on the "other half" of the process, i.e., on the "osmotic" type of dissipation of energy from the particle to its environment. (The VFT, then, gives relative probabilities for the respective cases, to which we shall return below.)

In fact, Equ. (2.1) already suggests to consider also contributions $-\delta\mathbf{p}$ in accordance with a particle's heat dissipation. Thus, returning to Equ. (2.19), and remembering the strict directionality of any heat flow, we can redefine this equation for the case of heat dissipation where $\Delta Q = Q(t) - Q(0) < 0$. Maintaining the heat flow as a positive quantity, i.e., in the sense of measuring the positive amount of heat dissipated into the environment, one therefore chooses the negative of the above expression, $-\Delta Q$, and inserts this into (2.19), to provide the osmotic velocity

$$\bar{\mathbf{u}} = -\mathbf{u} = D\frac{\nabla P}{P} = -\frac{1}{2\omega m}\nabla Q, \qquad (3.3)$$

and the osmotic current is correspondingly given by

$$\bar{\mathbf{J}} = P\bar{\mathbf{u}} = D\nabla P = -\frac{P}{2\omega m}\nabla Q. \qquad (3.4)$$

Then the corollary to Fick's second law becomes

$$\frac{\partial P}{\partial t} = -\nabla \cdot \bar{\mathbf{J}} = -D\nabla^2 P = \frac{1}{2\omega m}\left[\nabla P \cdot \nabla Q + P\nabla^2 Q\right]. \qquad (3.5)$$

With these ingredients at hand, let us now return to the expression (2.20) for the quantum potential $U$, and let us see how we can understand its thermodynamic meaning. To start, we study the simplest case, which is nevertheless very interesting. Certainly, one situation that is comparable for both the quantum and the classical descriptions is when we have to do with one free particle along a single path. Then, of course, both in the quantum and in the classical case, the quantum potential will vanish identically. (Remember at this point that in our thermodynamic ansatz, the quantum potential does have a "classical" meaning, too, as it actually was derived from purely classical physics.)

So, let us now consider $U = 0$, and then focus on the dynamics as we follow the behaviour of the osmotic velocity (3.3) that represents the heat dissipation from the particle into its environment. Firstly, we have from (2.20) that the thermodynamic corollary of a vanishing quantum potential reads as

$$\nabla^2 Q = \frac{1}{2\hbar\omega}(\nabla Q)^2. \tag{3.6}$$

Next, we insert (2.5) into (2.3) to give

$$P = P_0 e^{-\frac{\Delta Q}{\hbar\omega}}, \tag{3.7}$$

and thus also

$$\frac{\partial P}{\partial t} = -\frac{P}{\hbar\omega}\frac{\partial Q}{\partial t}. \tag{3.8}$$

Now, from the osmotic flux conservation (3.5) one has

$$\frac{\partial P}{\partial t} = \frac{P}{2\omega m}\left[\nabla^2 Q - \frac{(\nabla Q)^2}{\hbar\omega}\right]. \tag{3.9}$$

Thus, one obtains with (3.6), (3.8), and (3.9) that

$$\frac{1}{2}\left(\frac{\nabla Q}{\hbar\omega}\right)^2 = \frac{1}{D}\frac{1}{\hbar\omega}\frac{\partial Q}{\partial t}. \tag{3.10}$$

Insertion of (3.10) into (2.20) finally provides, with $\hbar\omega =$ constant and $\widehat{Q} := Q/\hbar\omega$,

$$U = -\frac{\hbar^2}{4m}\left[\nabla^2 \widehat{Q} - \frac{1}{D}\frac{\partial \widehat{Q}}{\partial t}\right], \tag{3.11}$$

or, generally,

$$U = -\frac{\hbar}{4\omega m}\left[\nabla^2 Q - \frac{1}{D}\frac{\partial Q}{\partial t}\right]. \tag{3.12}$$

As $U = 0$, one obtains

$$\nabla^2 Q - \frac{1}{D}\frac{\partial Q}{\partial t} = 0, \tag{3.13}$$

which is nothing but the *classical heat conduction equation*. In other words, *even for free particles*, both in the classical and in the quantum case, one can identify a *heat dissipation process emanating from the particle. A non-vanishing "quantum potential"*, then, is a means to describe the *spatial and temporal dependencies of the corresponding thermal flow in the case that the particle is not free*.

As to solutions to (3.13), let us consider some approximations at first. If, for example, we chose our temporal resolution such that, on average, the heat flow was constant, we would obtain from (3.13) a simple Laplace equation,

$$\nabla^2 Q = 0, \tag{3.14}$$

whose solutions are harmonic functions, as, in three dimensions, and with $r = \sqrt{x^2 + y^2 + z^2}$, given by $Q \propto \frac{1}{r}$, i.e., a singularity like a unit point charge at the origin. Further, if the temporal behaviour of $Q$ were not periodic (as we shall discuss lateron), but represented a heat generated only once and then dissipated, i.e., $Q \propto e^{-\omega t}$, then $\dot{Q} = -\omega Q = -Dk^2 Q$. Thus, with (3.13) one would then obtain the spatial Helmholtz equation

$$\left(\nabla^2 + k^2\right) Q = 0 \tag{3.15}$$

with periodic solutions $Q \propto e^{-i\mathbf{k}\cdot\mathbf{r}}$. Extending to the case of the inhomogeneous Helmholtz equation, consider the source as given by a delta function, i.e.,

$$\left(\nabla^2 + k^2\right) Q(\mathbf{x}) = \delta(\mathbf{x}). \tag{3.16}$$

For a unique solution one introduces the Sommerfeld radiation condition, for example, as a specification of the boundary conditions at infinity. Then $Q(\mathbf{x})$ is well known to be identical to a Green's function $G(\mathbf{x})$, which is given in three dimensions as

$$Q(\mathbf{x}) = G(\mathbf{x}) = \frac{e^{ik|\mathbf{x}|}}{4\pi|\mathbf{x}|}. \tag{3.17}$$

So, we see that the heat equation (3.13) provides, even for the case that $U = 0$, wave-like solutions $Q$, with the origin, or the particle's position, as source. In other words, one can identify solutions of the heat equation with Huygens-type waves.

Considering now the general case $U \neq 0$, we introduce in analogy to (1.2) an explicitly nonvanishing source term for the quantum potential (3.12), i.e.,

$$\nabla^2 Q - \frac{1}{D}\frac{\partial Q}{\partial t} = q(x)e^{i\omega t} \neq 0. \tag{3.18}$$

Depending on the presence of suitable (classical) potentials, or boundary conditions, respectively, one can often solve this equation via separation of variables. Thus, with the ansatz

$$Q = X(x)T(t), \quad \text{with} \quad T = e^{i\omega t}, \tag{3.19}$$

one has $\nabla^2(XT) = T\nabla^2 X = \frac{1}{D} X \frac{\partial}{\partial t} T + q(x)T$. With

$$q(x) := \alpha(x) X, \tag{3.20}$$

this becomes $T\nabla^2 X = \frac{1}{D} X \frac{\partial}{\partial t} T + \alpha XT$. Division by $(XT)$ then provides the constant

$$\frac{\nabla^2 X}{X} = \frac{\frac{\partial}{\partial t} T}{DT} + \alpha = -\lambda. \tag{3.21}$$

Considering, for example, the case of a particle entrapped in a box of length $L$ whose walls are infinitely high, one can introduce the Dirichlet boundary conditions, i.e.,

$$Q(0,t) = Q(L,t) = 0, \tag{3.22}$$

which provides the constant $\lambda$ as

$$\lambda = \frac{n^2\pi^2}{L^2} =: k_n^2. \tag{3.23}$$

With (3.21), normalization $\mathcal{N}$ and dimensionality preserving constant $C_Q$, one obtains

$$X = \mathcal{N} C_Q \sin\left(\frac{n\pi}{L} x\right), \qquad (3.24)$$

and, furthermore, with $T = e^{i\omega_n t}$ one has $\dfrac{i\omega_n}{D} + \alpha = -k_n^2$, and therefore

$$\alpha = -k_n^2 (1+i). \qquad (3.25)$$

With (3.20) one thus obtains

$$q(x) = -k_n^2 (1+i) \mathcal{N} C_Q \sin\left(\frac{n\pi}{L} x\right), \qquad (3.26)$$

and, with (3.19), (3.23), and $\tilde{Q} := Q/C_Q$,

$$\tilde{Q}(x,t) = \mathcal{N} \sin(k_n x) e^{i\omega_n t}. \qquad (3.27)$$

Note that, due to the Dirichlet boundary conditions, $e_n := \mathcal{N} \sin(k_n x)$ are eigenvectors of the Laplacian,

$$\nabla^2 e_n = -k_n^2 e_n, \qquad (3.28)$$

and

$$\langle e_n, e_m \rangle = \int e_n(x)\, e_m(x)\, dx = \begin{cases} 0 & m \neq n \\ 1 & m = n \end{cases}. \qquad (3.29)$$

This means that, for $m = n$, (3.29) can be interpreted as a probability density, with

$$\int_0^L P\, dx = \mathcal{N}^2 \int_0^L \sin^2(k_n x)\, dx = 1. \qquad (3.30)$$

The normalization in our example thus derives from (3.30) as

$$1 = \mathcal{N}^2 \int_0^L \frac{1}{2}\left[1 - \cos(2 k_n x)\right] dx = \mathcal{N}^2 \frac{L}{2},$$

and therefore

$$\mathcal{N} = \sqrt{\frac{2}{L}}. \qquad (3.31)$$

So, one obtains the result that the heat distribution in the box is given by

$$\tilde{Q}(x,t) = \sqrt{\frac{2}{L}} \sin(k_n x) e^{i\omega_n t}, \qquad (3.32)$$

with the probability density

$$P = \left|\widetilde{Q}(x,t)\right|^2. \tag{3.33}$$

Thus, the classical state (3.32) is shown to be identical with the quantum mechanical one [5], and one obtains with (3.33) and (2.16) the quantum potential as

$$U = \frac{\hbar^2 k_n^2}{2m}. \tag{3.34}$$

Note that with (3.26), one derives from (3.18) a sort of "eigenvalue equation",

$$\nabla^2 \widetilde{Q} - \frac{1}{D}\frac{\partial \widetilde{Q}}{\partial t} = q(x)e^{i\omega t} = -(1+i)k_n^2 \widetilde{Q}, \tag{3.35}$$

with the unique solutions for $\widetilde{Q}$ in our example given by (3.32). However, taking account of the correct coefficients from (3.11), we do not (yet) obtain from (3.35) the correct expression (3.34) for $U$. To solve this problem we must remember that in deriving (3.11) and (3.12), we solely focused on the osmotic current of heat dissipation. Doing the same calculation for the usual forward current, and then adding both to obtain the total current, should provide the correct answer, then. In fact, this can easily be seen, as the forward current only provides a minus sign in front of the r.h.s. of (3.5) and thus also of (3.10). This is just another way of saying that the two currents can be considered as "forward" and "osmotic" fluxes, distinguished merely by the sign in their temporal behaviours. It then follows that the total quantum potential, integrated over periods of time $t \sim n/\omega$ long enough so that it is characterized by the energy throughput of $n$ total currents, or by equal weights of $n$ "absorption" and $n$ "dissipation" currents, respectively, is with (3.12) given by

$$\overline{U} = \frac{1}{n}\left(n\overline{U}_{\text{forward}} + n\overline{U}_{\text{osmotic}}\right) = -\frac{\hbar^2}{4m}\frac{1}{\hbar\omega}\left[2\nabla^2 Q + \frac{1}{D}\frac{\partial Q}{\partial t} - \frac{1}{D}\frac{\partial Q}{\partial t}\right] = -\frac{\hbar^2}{2m}\frac{\nabla^2 Q}{\hbar\omega}. \tag{3.36}$$

As in the case of the particle-in-a-box the total energy is completely thermalized [5], one has $\hbar\omega = Q$, and thus one obtains with (3.27) the correct value (3.34) for $U = \overline{U}$. Note that heuristically one can arrive at the result (3.36) also as follows. As the vacuum fluctuation theorem [3] describes a statistical dominance of heat dissipation over heat absorption, one has upon averaging over many processes of energy throughput that the average $\overline{P} = P_0 e^{\delta Q/\hbar\omega}$ upon insertion into the definition of the quantum potential, $U = -\frac{\hbar^2}{2m}\nabla^2\sqrt{P}/\sqrt{P}$, provides also $U = -\frac{\hbar^2}{2m}\frac{\nabla^2 Q}{\hbar\omega}$. This "shortcut" type of derivation, however, would not reveal the deeper reason for the obtained result, that is, Equ. (3.35), which has a simple explanation, to be given now.

Note, firstly, that an equation of the form (3.35) also holds for the one-particle case of two intersecting paths [5], such that even for "free" particles one can have a non-vanishing quantum potential. However, most importantly, one can observe that by applying a temporal Fourier transformation on (3.35) and introducing the complex diffusion wave number $\kappa(x,\omega) := \sqrt{\frac{i\omega}{D}}$, one obtains a Helmholtz-type pseudo-wave equation:

$$\nabla^2 \widetilde{Q}(x,\omega) - \kappa^2 \widetilde{Q}(x,\omega) = Q(x,\omega). \qquad (3.37)$$

Recalling Equ. (1.4) from the introduction, we see that (3.37), along with the identical definition of $\kappa$, is exactly the *defining equation for a thermal-wave-field* and thus describes the spatio-temporal behaviour of *diffusion waves*. Moreover, with the thermalized version of $U$, (3.36), one can now generally rewrite the usual modified (i.e., by the "quantum potential") Hamilton-Jacobi equation (2.15), using (1.5) and $D = \hbar/2m$, as

$$\frac{\partial S}{\partial t} + \sum_{i=1}^{n} \frac{(\nabla_i S)^2}{2m_i} + V - \sum_{i=1}^{n} \left(\frac{L(\omega_i)}{2}\right)^2 \nabla_i^2 Q = 0. \tag{3.38}$$

Thus, the equations of quantum motion [3, 8] can be written as

$$m_i \frac{d\mathbf{v}_i}{dt} = -\nabla_i (V + U) = -\nabla_i V + \sum_{i=1}^{n} \left(\frac{L(\omega_i)}{2}\right)^2 \nabla_i \left(\nabla_i^2 Q\right). \tag{3.39}$$

It is therefore possible that, at least for simple solutions $Q$, the calculation of quantum trajectories becomes simplified as well.

## 4. The quantum potential as a representation of the accumulation-depletion law for diffusion-wave fields

Considering thermal waves as a typical application of the theory of parabolic diffusion-wave fields (DWF), their most conspicuous properties relate to their distinction from hyperbolic waves. Mandelis et al. [6] have shown in detail that if a point source generates thermal waves at angular modulation frequency $\omega$, the diffusion of these waves is generally not in the direction of the wave vector. Instead, the flux/current is solely driven by the temperature/heat gradients, viz. (3.4), and the spatial distribution of the DWF is proportional to the Green function related to the source coordinate. In fact, Mandelis' book on DWF [7] contains essentially a large list of examples of Green function solutions to various problems regarding interactions at interfaces. In [6] and [7], the authors introduce a "thermal-wave interface-interaction coefficient" for solutions of the wave equation, where each term multiplying this coefficient is "associated with the value of the flux *after* the interaction of the thermal wave with the interface". At least three flux vectors can be discerned at any point of an interface: incident ($\mathbf{J}_i$), interface-interacted ($\mathbf{J}_r$), and transmitted ($\mathbf{J}_t$). For the simple example discussed in [6], the authors show that the currents are subject to simple conservation laws. Depending on the sign of (the real part of) above-mentioned coefficient, these conservation laws either describe *thermal-wave energy accumulation* within some spatial domain, or *thermal-wave energy depletion*, respectively. It is only for the case when the flux conservation law reduces to $\mathbf{J}_r = -\mathbf{J}_i$ that one can speak of a "perfect accumulation condition". This case reproduces the reflection law of conventional, hyperbolic wave fields, and it is in accordance with the observation that the greater the transport property discontinuity across the interface, the more reflection-like the behaviour is. In that case, then, normal incidence on perfect mirrors does produce standing waves, just like the ones of the particle-in-a-box example discussed above.

So, due to Fick's laws and the corresponding unidirectionality of thermal flows, one can point out the main characteristic of DWF as given by the field-gradient-driven accumulation-depletion laws. As we have in the preceding Section identified the quantum potential part of the generalized Hamilton-Jacobi equation with the presence of DWF, we can now consider some examples of various forms of the quantum potential, or Bohmian trajectories, respectively, and try to understand them in terms of the presence of DWF.

    *a) Gaussian wave-packet and potential barrier*

We firstly consider a Gaussian wave-packet whose particle trajectories are partly reflected and partly tunnelling through a potential barrier. Fig. 1 shows the temporal behavior (vertical) in one spatial dimension (horizontal) of the wave-packet with 15 sample trajectories (yellow) and the bluish-white background indicating the probability density distribution.

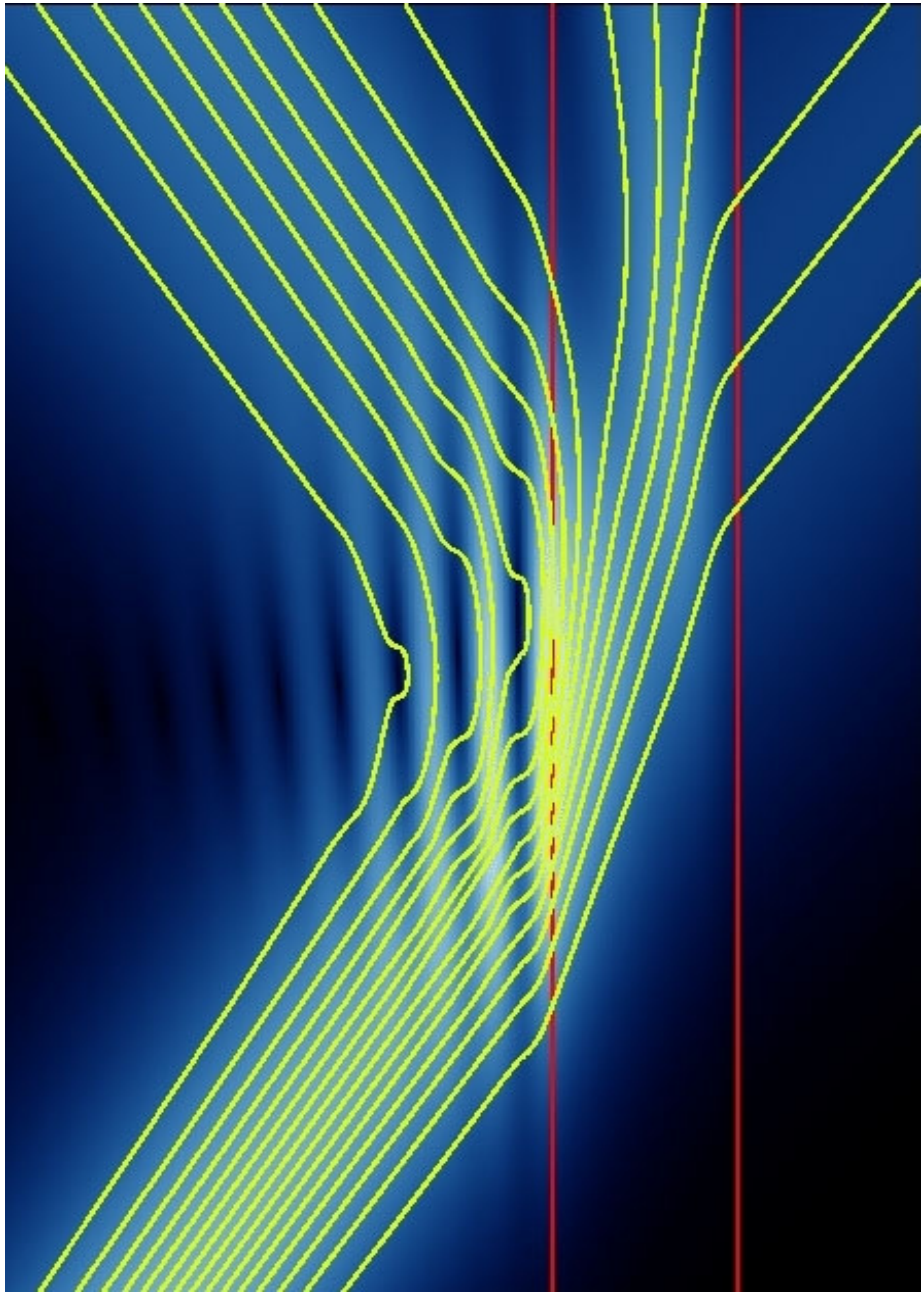

Fig.1: Gaussian wave-packet interacting with a potential barrier (red lines), shown in one spatial (horizontal) and one temporal (vertical) dimension. (After Dewdney and Hiley [8].) The probability density distribution (in blue-to-white) clearly indicates the presence of accumulation and depletion zones (i.e., the "stripes" roughly parallel to the barrier) of the diffusion-wave field.

Note that the probability density distribution comes in "stripes" roughly parallel to the barrier. These stripes represent accumulation and depletion zones of the DWF, the totality of which is just the "thermodynamic version" of the quantum potential. Note particularly that the roughly sinusoidal distribution of stripes represents the result of the interface-interacted thermal flux, whose outward vector is approximately normal to the barrier wall. Thus, for example, in the (bright) accumulation zones, many of the incoming particles experience a strong "force" away from the barrier due to the permanent buffeting from the thermal field, whereas the (dark) depletion regions "release" the particles from the impinging thermal wave again such that they can roughly re-obtain the directions of their original momenta. As a consequence, some of the particles are "bouncing off" the thermal wave field even before they reach the barrier.

*b) Free particle evolution with two or more paths*

As mentioned, the quantum potential can be different from zero also in the free particle case. Thus, even without the presence of an external potential, the intersecting region of particle paths gives rise to complex behaviour.

In Fig. 2a, two wave packets of a single particle cross each other and are again seen to produce (vertical) "stripes" due to alternating high and low probability densities. The latter, in turn, are clearly representations of accumulation and depletion zones of the involved DWF. Similarly, in Fig. 2b the stripes appear in many occasions, when six (differently weighted) wave packets interact.

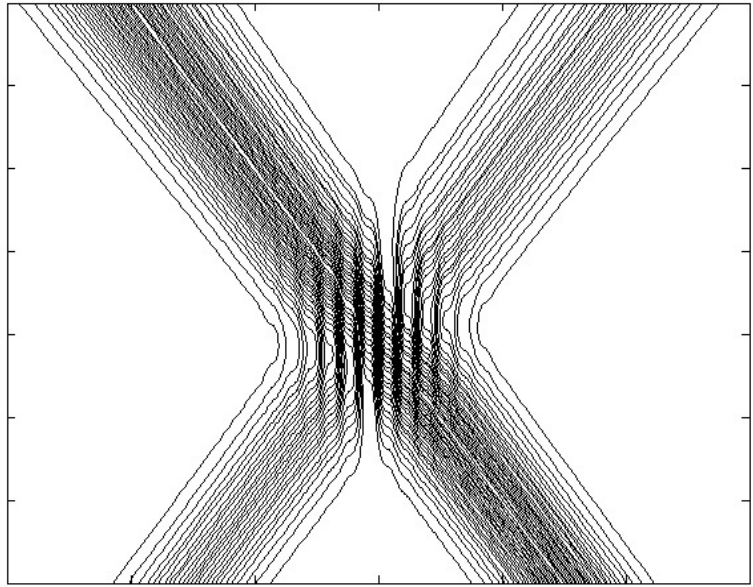

Fig. 2a: Spacetime diagram of two possible paths of a particle crossing each other. Note again the vertical "stripes", representing alternating high and low probability densities, or the corresponding accumulation and depletion zones of DWF, respectively. (From [9].)

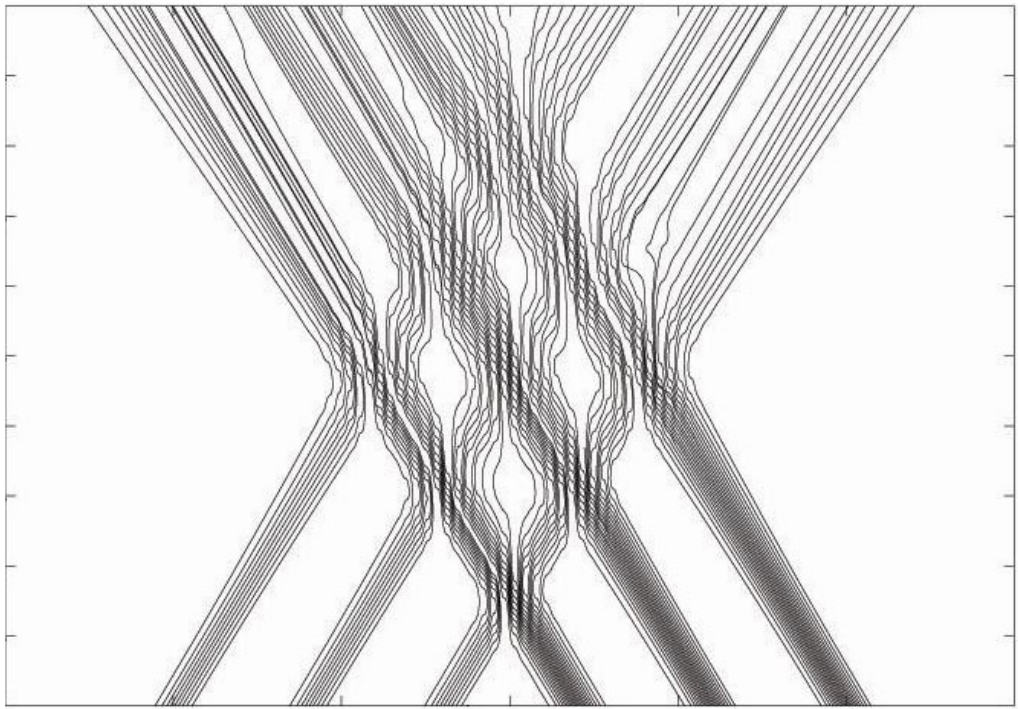

Fig. 2b: Six paths of a single particle, three from the left, and three (differently weighted ones) from the right. (From S. Kreidl [10].)

*c) Diffraction at the double slit*

Figs. 3a and 3b show two well-known pictures related to the de Broglie-Bohm interpretation. Referring to diffraction at a double slit, one Gaussian wave packet emerges from each slit, thereby producing the quantum potential (a) and the corresponding trajectories (b). Note that the quantum potential is also in this example characterized by the appearance of "stripes", or "deep valleys" and "high-level plateaus", respectively. As can be seen from the corresponding behaviour of the trajectories, the particles are accelerated in the "valleys" and move unhindered on the plateaus. Even "stranger" still is the role of the central axis: the particle trajectories from each side are "bouncing off" said symmetry line such that no particle from the left slit can enter the right region, and vice versa. This bouncing off has been criticized as "unphysical" or even "surrealistic", because apparently, on a more macroscopic scale taking into account the positions of the two slits, momentum conservation would be severely violated. However, if we look at the situation in terms of the presence of DWF, that criticism completely dissolves, and one can show that an overall momentum conservation is given.

Thus, we explain Figs. 3a and 3b as follows. Imagine particles originating from the slits and emitting initially spherically symmetrical thermal waves from the origin of their respective positions. Then let the whole field evolve and consider what happens after completion of the interaction to the whole thermal field. Of course, the "heat" will dissipate far away from the slit system, making the quantum potential effectively flat on its outer edges. However, in between the two slits, the heat waves accumulate, thus creating the "mountain" of heat (i.e., a kinetic energy reservoir) in between the slits, such that the particles coming from the slits later on will immediately experience strong changes of momentum in the direction radiating away from the "mountain". Particularly, the central axis between the two slits will be such that, to the left and to the right of it, the particles will take up such high momentum from the "thermal mountain" that a crossing into the respective other half beyond the symmetry axis is made impossible. Thus, we see that, just like in the previous examples, also in the case of a double slit, "stripes" emerge which represent nothing other than accumulation and depletion zones of the DWF. The bouncing-off from accumulation zones around the central axis thus is very similar to the bouncing-off from the accumulation zones of Fig. 1 even before particles reach the barrier wall. Note in particular, moreover, that the quantum potential represents the thermal wave field in its totality, as the latter is already there across the whole experimental setup, i.e., due to the infinitely fast propagation of the DWF. This explains the peculiar appearance of the quantum potential as well as the unusual behaviour of particle motion. In any case, however, the latter can be imagined under the premises of complete conservation of momentum, as long as the momentum contributions $\delta\mathbf{p}$ of the thermal wave field are taken into account, i.e., in addition to the particle's momentum $\mathbf{p}$.

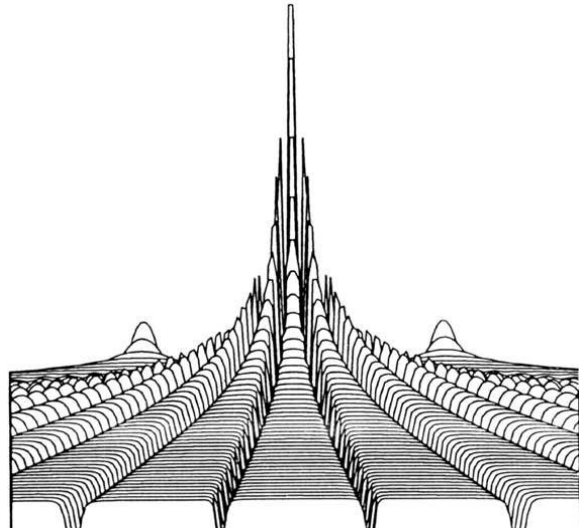

Fig. 3a: Quantum potential for two Gaussian slits, from [12]. Both the "mountain" in between the two slit regions and the "stripes" of alternating valleys and plateaus, respectively, are manifestations of the structured reservoir of heat (kinetic energy) due to the presence of accumulation and depletion zones of the nonlocal diffusion-wave field.

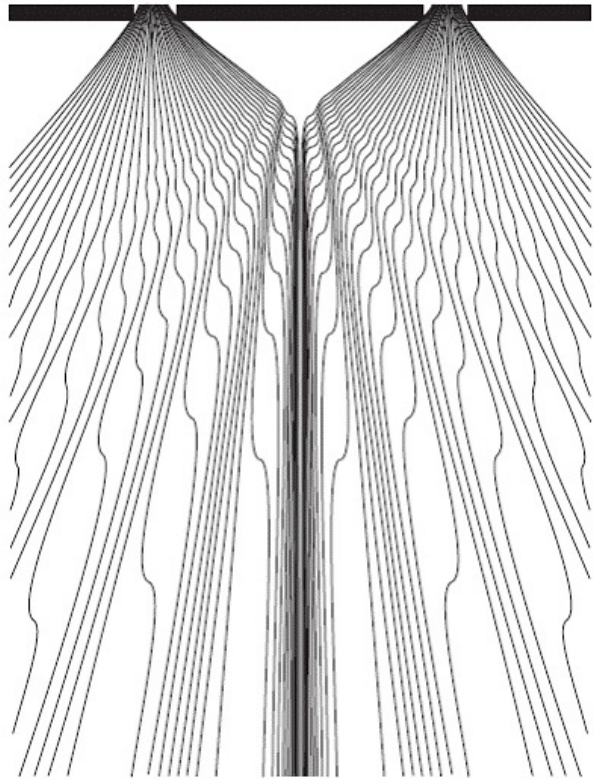

Fig. 3b: Trajectories according to Fig. 3a), after ref. [12]. Note the drastic changes in the directions of the momenta of particles close to the area of the central "mountain" of kinetic energy, pushing the particles onto the accumulated plateaus of Fig. 3a, whereas in the depleted valleys the particles are temporarily "released" to follow their original momentum vectors.

   *d) Two wave packets impinging on a barrier with a phase difference*

Finally, in Fig.4, the effective (i.e., classical plus quantum) potential is shown for two incident packets from two sides with a phase difference of $\Delta\Phi = \pi/2$. Here, we can see the "stripes" very clearly, both in the representation of the quantum potential (a) and in the particle density distribution indicated by the bunching of the particle trajectories (b). We note particularly that, although in general diffusion waves do not exhibit wave fronts, interference effects, and the like, they nevertheless do take account of spatially correlated phase lags, which are essentially determined by the diffusion length. [7] This, however, guarantees that also nonlocal correlations, testable through variations of phase lags, can in principle be accounted for by the physics of DWF.

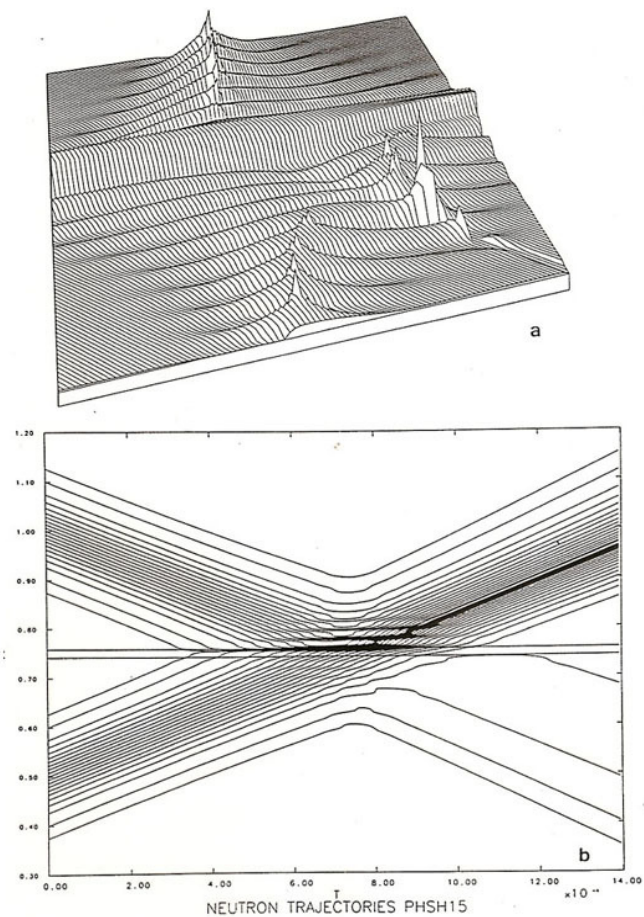

Fig.4: (a) Effective potential for two incident wave packets impinging on a barrier, with phase difference $\Delta\Phi = \pi/2$. (b) Trajectories for the situation of (a). (From C. Dewdney [15].) The accumulation and depletion zones are clearly discernible in both representations.

## 5. Conclusions and outlook

According to the vacuum fluctuation theorem [3], there is a non-zero probability for even the particles of quantum theory to dissipate a finite amount of heat $Q$ into the surrounding environment. Both in possible equilibrium models and in the nonequilibrium approach advocated in [3], the heat capacity of the thermal reservoir of the surrounding vacuum is such that the equilibrium-type Maxwell-Boltzmann distribution $P$ can safely be assumed, providing essentially $\ln P \propto -Q$. Inserting the corresponding exact relation into the defining equation for the quantum potential $U$, together with the constraints of flux conservation as given by Fick's law (i.e., essentially substituting $-\delta Q$ by $+\delta Q$), enables one to rewrite the expression for $U$ completely in terms of a sub-quantum thermodynamics. For vanishing $U$, the result is exactly identical to the classical heat equation, whereas for $U \neq 0$ one obtains a parabolic wave equation whose solutions are given in terms of classical diffusion-wave fields (DWF).

In this way, the "form" of the quantum potential, as given by $-\nabla^2 R/R = -1/2\left(\nabla^2 \ln P\right)$, translates essentially into a Helmholtz-type dependence $-\nabla^2 Q$ of a thermal energy $Q$ distributed "nonlocally" throughout an experimental apparatus, for example. As the corresponding wave equation is a parabolic one, one must generally expect the thermal waves' behaviour to be radically different from that of ordinary hyperbolic ones. In fact, we have discussed several examples of forms of the quantum potential which clearly exhibit the departure from ordinary wave behaviour in terms of the appearance of accumulation and depletion zones. Thus, the "strange" form of the quantum potential as well as the (only seemingly "surrealistic") Bohmian trajectories can potentially be fully understood with the aid of the physics of DWF.

We note particularly the nonlocal features of DWF. As their "propagation speed" is infinite, one must imagine the following scenario for, e.g., an experiment in neutron interferometry. With a prepared neutron source in a reactor, one immediately has a thermal field in the "vacuum" that nonlocally links the neutron oven, the apparatus (including, e.g., a Mach-Zehnder interferometer), and the detectors. The (typical) Gaussians used to describe the initial quantum mechanical particle distributions thus also contribute in their totality to the form of the heat distribution in the overall system, no matter which particle actually is on its way through the interferometer. In this way, all "potential" paths are implicitly present throughout the experiment (i.e., under constant boundary conditions) in that the corresponding thermal field exists no matter where the particle actually is. That this can be assumed is, of course, solely due to the fact of the infinite propagation of DWF.

It is clear that in this way a promising perspective for a deeper understanding of quantum mechanical nonlocality is opened up. In this regard, one of the most exciting questions to be tackled in the future will be how all of this can be described in relativistic terms. One may well envisage a theory of relativistic DWF as a new opportunity to reconcile quantum theory and relativity.

**References**

[1]     Bohm, D. and Hiley, B. J., *The Undivided Universe*. Routledge, London (1993)

[2]     Mandelis, A., Diffusion Waves and their Uses, *Phys. Today* 53, 29 (2000)

[3]     Grössing, G., The vacuum fluctuation theorem: Exact Schrödinger equation via Nonequilibrium Thermodynamics, *Phys. Lett. A* 372, 4556 (2008); see also http://arxiv.org/abs/0711.4954 .

[4]     Evans, D. J. , Searles, D. J., The Fluctuation Theorem, *Adv. Phys.* 51, 1529 (2002)

[5]     Grössing, G., Diffusion Waves in Sub-Quantum Thermodynamics: Resolution of Einstein's 'Particle-in-a-box' Objection, http://arxiv.org/abs/0806.4462 , to be published (2008)

[6]     Mandelis, A., Nicolaides, L., and Chen, Y., Structure and the Reflectionless/ Refractionless Nature of Parabolic Diffsion-Wave Fields, *Phys. Rev. Lett.* 87, 020801 (2001)

[7]     Mandelis, A., *Diffusion-Wave Fields.* Springer, New York (2001)

[8]     Dewdney, C. and Hiley B. J., A Quantum Potential Description of One-Dimensional Time-Dependent Scattering from Square Barriers and Square Wells, *Found. Phys*. 12, 27 (1982)

[9]     From the Bohmian mechanics group at the University of Innsbruck, http://bohm-mechanics.uibk.ac.at/ .

[10]    Adapted from Kreidl, S., Grübl, G., and Embacher, H. G., *Bohmian arrival time without trajectories*, *J. Phys.* A: Math. Gen. **36,** 8851 (2003); with kind permission from Sabine Kreidl, Univ. Innsbruck.

[11]    Phillipidis, C., Dewdney, C., and Hiley, B.J., Quantum Interference and the Quantum Potential, *Nuovo Cimento* 52B, 15 (1979)

[12]    Dewdney, C., The quantum potential approach to neutron interferometry, *Physica* B 151, 160 (1988)